\documentclass{article}
\usepackage{graphicx}  
\usepackage{amsmath}   
\usepackage[compress]{cite}
\usepackage{amssymb}   
\usepackage{bm} 
\usepackage{dcolumn}
\usepackage{color}
\usepackage{mathrsfs}
\usepackage{amsfonts}
\usepackage{varioref}
\usepackage{textcomp}

\RequirePackage[colorlinks,citecolor=blue,urlcolor=magenta,linkcolor=blue]{hyperref}
\allowdisplaybreaks
\labelformat{section}{Section #1} 
\labelformat{subsection}{Section #1} 
\labelformat{subsubsection}{Section #1}
\labelformat{subsubsubsection}{Section #1}
\labelformat{equation}{Eq.~(#1)} 
\labelformat{figure}{Fig.~#1} 
\labelformat{subfigure}{Fig.~\thefigure#1} 
\labelformat{table}{Table~#1} 
\labelformat{appendix}{Appendix #1}
\newcounter{mnotecount}[section]

\renewcommand{\themnotecount}{\thesection.\arabic{mnotecount}}

\newcommand{\mnotex}[1]
{\protect{\stepcounter{mnotecount}}$^{\mbox{\footnotesize
$
\bullet$\themnotecount}}$ \marginpar{
\raggedright\small\em
$\!\!\!\!\!\!\,\bullet$\themnotecount: #1} }

\begin{document}
\title{Looking for static interior solutions of  Buchdahl star with $p_r=0, p_t=k\rho$ in general relativity and pure Lovelock theories}
\author{Shauvik Biswas
\footnote{shauvikbiswas2014@gmail.com}$~^{1}$
and Chiranjeeb Singha\footnote{chiranjeeb.singha@saha.ac.in}$~^{2}$\\
$^{1}$\small{School of Physical Sciences, Indian Association for the Cultivation of Science, Kolkata-700032, India}\\
$^{2}$\small{Theory Division, Saha Institute of Nuclear Physics, Kolkata 700064, India}}
\date{\today}
\maketitle

\begin{abstract}
We find static fluid solutions of Einstein and pure Lovelock equations with $p_r=0$, $p_t=k\rho$, which could be possible models for the interior of a Buchdahl-like star. 
Buchdahl star is a limiting stellar configuration without a horizon whose formation does not need any exotic matter.   
\end{abstract}

\section{Introduction} 

It is well known that for solving the gravitational equation for the fluid interior of a static stellar object, one has either to prescribe an equation of state, that is, a relation between pressure and density, or one of the metric functions to close the system of equations. In a recent paper\cite{dad23}, it is inferred from very general considerations that Buchdahl-like interior may possibly have an equation of state,  $p=\frac{1}{2}\rho$. This, however, is not sustainable on simple physical consideration that at the boundary $p=0$ implying sound velocity being zero, which contradicts $v_s^2 = dp/d\rho =1/2 \neq 0$. One possible way out is to assume $p_r=0$, which takes care of the boundary condition and has $2p_t = \rho$. This is precisely our motivation for finding a static fluid sphere with $p_r = 0$ and $p_t = k\rho$ as possible models of Buchdahl-like interior. \\

Recently the characteristics of spacetime near the event horizon have been probed by the gravitational wave measurements \cite{LIGOScientific:2016aoc, LIGOScientific:2021sio, LIGOScientific:2016lio, LIGOScientific:2017bnn, LIGOScientific:2016vlm, LIGOScientific:2017vwq, LIGOScientific:2018mvr, LIGOScientific:2020ibl} and the observation of black hole shadow \cite{EventHorizonTelescope:2019ths, EventHorizonTelescope:2019dse, EventHorizonTelescope:2019ggy, Gralla:2019xty}. It is claimed that there is room for alternatives \cite{Cardoso:2019rvt, Biswas:2022wah, Dey:2020pth, Maggio:2021ans, Carballo-Rubio:2018jzw, Vagnozzi:2019apd, Cunha:2018acu, Banerjee:2019nnj, Mishra:2019trb} which can mimic a general relativistic black hole. In particular, within the current experimental errors, the gravitational wave signal will very much
look like that of a black hole, even if stellar structures exist within the region between the black hole horizon and
the photon sphere. Thus, studying the physical properties of such stellar structures is very important. \\

Buchdahl star is a limiting stellar configuration without a horizon but having a photon sphere. Using the finiteness of central pressure, one can obtain the Buchdahl bound without solving the T-O-V (Tolman–Oppenheimer–Volkoff) equations for a uniform stellar interior\cite{padmanabhan2010gravitation}. In general, it is defined by potential, $\phi (R)=M/R=4/9$ \cite{Dad22}. For a purely geometric version of the Buchdahl star characterization, see \cite{Dadhich:2023csk}. Buchdahl star shares almost all the property of the black hole \cite{Chak22, Shaymatov:2022ako, Shaymatov:2022hvs}. The compactness limit of Buchdahl star has been considered in extensive literature like the inclusion of $\Lambda$ \cite{Mak:2001eb, article, Andreasson:2012dj, Stuchlik:2000gey}, different conditions than Buchdahl's \cite{Karageorgis:2007cy, Andreasson:2007pj}, brane-world gravity \cite{Germani:2001du, Garcia-Aspeitia:2014pna}, modified gravity theories including Lovelock gravity and higher dimensions \cite{Goswami:2015dma, PoncedeLeon:2000pj, Wright:2015yda, 2009GReGr..41..453Z,  Dadhich:2016fku, Feng:2018jrh}. One can obtain the limit on the maximum mass of the Buchdahl-star by appealing to the dominant energy condition where the sound velocity is subliminal \cite{PhysRevD.65.124028, Fujisawa:2015nda}. The Buchdahl bound is defined as an overriding state. This bound is obtained under very general conditions, while more compact distributions are allowed under certain circumstances and conditions. \\


Here we consider a perfect fluid as the source of the matter for getting the possible models of Buchdahl interior. In \ref{Einstein}, we find a static fluid solution of the Einstein equation with $p_r=0, p_t=k\rho$. We show that for $k=2$, we get the possible models of Buchdahl's interior in Einstein's gravity. In \ref{Lovelock}, we find the static fluid solutions of the Lovelock equation with $p_r=0$, $p_t=k\rho$.
We show that for $k=1/2$ and Lovelock order$N=3$, we get the possible models of the Buchdahl interior in pure Lovelock gravity. Although, for pure Lovelock theories, higher curvature terms are present in the Lagrangian, the field equations are still second-order \cite{Gannouji:2019gnb, Dadhich:2015nua, Dadhich:2012cv, Gannouji:2013eka, Dadhich:2015ivt, Padmanabhan:2013xyr, Dadhich:2015lra}. For these theories, in the Riemann curvature tensors, we define the Roman capital indices for the $(d-2)$ angular coordinates \footnote{We have used $c=G=1$ throughout the paper.}.


\section{Static interior solutions in general relativity}\label{Einstein}

Here we consider a spherically symmetric object whose exterior geometry is governed by the Schwarzschild metric. We want to explore the interior solution of the object. We consider the interior metric in the following form,
\begin{equation}
  ds^2=- A (t,r) dt^2+ B (t,r) dr^2+r^2 d \Omega^2~.\label{eqn1}  
\end{equation}
We also consider that the spacetime is static, so we chose $A (t,r)=A (r)$ and $B (t,r)= B(r)$. Now we compute the field equation. The left-hand side of the Einstein equation for this spacetime is given below,
\begin{eqnarray}
G^{t}{}_{t}&=&-\frac{B'}{r B^2}-\frac{1}{r^2}+\frac{1}{r^2 B}~,\nonumber\\
G^{r}{}_{r}&=&\frac{A'}{r A B}-\frac{1}{r^2}+\frac{1}{r^2 B}~,\nonumber\\
G^{\theta}{}_{\theta}=G^{\phi}{}_{\phi}&=& \frac{A'}{2 r A B}-\frac{B'}{2 r B^2}+\frac{A''}{2 A B}-\frac{A'}{4 A B}\left(\frac{A'}{A}+\frac{B'}{B}\right)~. 
\end{eqnarray}
Here `prime' denotes the derivative with respect to the radial coordinate $r$. Here we consider the perfect fluid as the source of the matter. The stress-energy tensor for the perfect fluid is given by $T^{\mu}{}_{\nu}= dig~(-\rho, P_{r}, P_{t}, P_{t})$. Then Einstein equation $G^{\mu}{}_{\nu} = 8 \pi T^{\mu}{}_{\nu}$ becomes,
\begin{eqnarray}
    -\frac{B'}{r B^2}- \frac{1}{r^2}+\frac{1}{r^2 B}= - 8 \pi \rho~,\label{eqn3}\\
    \frac{A'}{r A B}- \frac{1}{r^2}+\frac{1}{r^2 B}= 8\pi P_{r}~,\label{eqn4}\\
    \frac{A''}{2 A B}- \frac{A'}{4 A B}\left(\frac{A'}{A}+\frac{B'}{B}\right)+\frac{A'}{2 r A B}-\frac{B'}{2 r B^2}= 8 \pi P_{t}~.\label{eqn5}
\end{eqnarray}
One can also write \ref{eqn4} as,
\begin{equation}\label{eqn6}
    \frac{d}{d r}\left[r \left(1-\frac{1}{B}\right)\right]= 8 \pi \rho r ^2~.
\end{equation}
On the other hand, from the conservation of the matter energy-momentum tensor, \emph{i.e.,} $\nabla_\mu T^{\mu}{}_{\nu}=0$, we obtain
\begin{equation}\label{eqn7}
    P'_{r}=\frac{2}{r}\left(P_{t}-P_{r}\right)-\frac{A'}{2 A}\left(\rho+P_{r}\right)~.
\end{equation}
Here we are looking for a solution for the interior where $P_{r}=0$ and $P_{t}=k \rho$. Inserting $P_{r}=0$ and $P_{t}=k \rho$ in the \ref{eqn7}, one can immediately get, 
\begin{eqnarray}\label{eqn8}
\frac{A'}{A}=\frac{4 k}{r}~,
\end{eqnarray}
which implies that $A =A_0 (r/M)^{4 k}$. Here $A_{0}$ is the integration constant. We also get the solution of $B$ from \ref{eqn5} by setting $P_{r}=0$ and using the above expression of $A$. The solution of $B$ is then given by,
\begin{eqnarray}
    \frac{A'}{r A B}- \frac{1}{r^2}+\frac{1}{r^2 B}= 0\\
    \Rightarrow \frac{1}{r B} \frac{4 k}{r} -\frac{1}{r^2}+ \frac{1}{r^2 B}=0\\
    \Rightarrow B= 4 k+1~. \label{eqn11}
    \end{eqnarray}
Now putting the above expression of $B$ in \ref{eqn6}, we get,
\begin{eqnarray}
&\frac{d}{d r}\left[r \left(1-\frac{1}{4 k+1}\right)\right]= 8 \pi \rho r ^2~.\nonumber\\
&\Rightarrow \rho= \frac{4 k}{4k +1} \frac{1}{8 \pi r^2}~.\nonumber\\
&\Rightarrow \rho \sim \frac{1}{r^2}~.
\end{eqnarray}

We have shown that density $(\rho)$ is inversely proportional to $r^2$. Again, from \ref{eqn8}, one can easily calculate $A'$ and $A''$. Using the expression of $A'= \left(4 k A_{0} r^{4 k-1}\right)/M^{4k}$, $A''=\left(4 k (k-1) A_{0} r^{4 k-2}\right)/M^{4k}$ in \ref{eqn5}, we get $P_{t}=\frac{4 k^2}{4 k +1}\frac{1}{8 \pi r^2} \sim \frac{1}{r^2}$. Thus, $P_{t}$ is also inversely proportional to $r^2$. 

Here we consider the exterior spacetime as the Schwarzschild spacetime. The metric for exterior spacetime is given by,
\begin{equation}\label{Schwarzschild}
d s^2= \left(1- \frac{2 M}{r}\right) dt ^2+ \left(1- \frac{2 M}{r}\right)^{-1} dr^2+ r^2 d \Omega^2~,
\end{equation}
where $M$ is the ADM mass. We consider the interior and exterior solutions to match each other at some radius $R$. Then from \ref{eqn1}, \ref{eqn11}, \ref{Schwarzschild}  and the matching condition, we get,
\begin{eqnarray}
    B=4 k +1&=& \frac{1}{1-\frac{2 M}{R}}\nonumber\\
    \Rightarrow \frac{ M}{R}&=& \frac{2 k}{4k +1}~.
\end{eqnarray}
Also, we arrive at the same conclusion by considering $A'$ should be continuous at some radius $R$. Then from the \ref{eqn1}, \ref{eqn8}, \ref{Schwarzschild} and the matching condition, we get,
\begin{eqnarray}
    \frac{A'}{A}&=& \frac{2 M/R^2}{(1-2 M/R)}\nonumber\\
    \Rightarrow \frac{4 k}{R}&=&\frac{2 M/R^2}{(1-2 M/R)}\nonumber\\
    \Rightarrow \frac{ M}{R}&=& \frac{2 k}{4k +1}~.
\end{eqnarray}
For $k=1/2$, we get $M/R= 1/3$. For $k=1$, we get $M/R= 2/5$. For $k=2$, we get the potential for the Buchdahl star, where $M/R = 4/9$. For $k=3$, we get $M/R=6/13$. Thus, $p_r = 0$ and $p_t = 2 \rho$, we get the possible models of Buchdahl's interior in Einstein's gravity. 


\section{Static interior solutions in pure Lovelock theories}\label{Lovelock}

Here we do the same calculation for the pure Lovelock gravity in d-dimensional spacetime of Lovelock order N. We consider the interior metric of the object in the following form,
\begin{equation}\label{eqn161}
d s^2= - A (r) dt^2+ B(r) dr^2 +r^2 d \Omega^2_{d-2}.
\end{equation}
Here we also consider the metric is static and spherically symmetric. We take the source of the matter as a perfect fluid. So the energy-momentum tensor can be written as $T^{\mu}{}_{\nu}=(-\rho,P_{r},P_{t},P_{t})$. The corresponding field equations are given by \cite{Dadhich:2016fku, Chakraborty:2021dmu},
\begin{eqnarray}
    8 \pi P_{r}= \frac{\left(1-\frac{1}{B}\right)^{N-1}}{2^{N-1}r^{2N}}\left[rN \frac{A'}{AB}-(d-2N-1)\left(1-\frac{1}{B}\right)\right]~,\label{eqn16}\\
    8 \pi \rho= \frac{\left(1-\frac{1}{B}\right)^{N-1}}{2^{N-1}r^{2N}}\left[rN \frac{B'}{B^2}+(d-2N-1)\left(1-\frac{1}{B}\right)\right]~.\label{eqn17}
\end{eqnarray}
In the context of d-dimensional spacetime, the conservation of the energy-momentum tensor is given below \cite{Chakraborty:2021dmu},
\begin{equation}
P_{r}'+ \frac{A'}{2 A}(\rho+P_{r})+\left(\frac{d-2}{r}\right)(P_{r}-P_{T})=0~.\label{eqn18}
\end{equation}
Here also, we are looking for a solution for the interior where $P_{r}=0$ and $P_{t}= k \rho$.  Inserting $P_{r}=0$ and $P_{t}=k \rho$ in the \ref{eqn18}, one can immediately get, 
\begin{eqnarray}\label{eqn19}
\frac{A'}{A}&=&\frac{2(d-2)k}{r}\nonumber\\
\Rightarrow A &=& A_{0} (r/M)^{2(d-2)k}~,
\end{eqnarray}
where $A_{0}$ is the integration constant. Substituting the expressions $A$ from \ref{eqn19} and setting $P_{r}=0$ in the \ref{eqn16}, we get,
\begin{equation}\label{eqn20}
B=\frac{\left[2 (d-2) k N+ (d-2 N-1)\right]}{(d-2 N -1)}~.
\end{equation}
Again from \ref{eqn17}, using \ref{eqn19} and \ref{eqn20}, we get $\rho \sim \frac{1}{r^{2N}}$. As $P_{t}=k \rho$ so we get $P_{t} \sim \frac{1}{r^{2 N}}$. We have shown that for this case, density $(\rho)$ and $P_{t}$ are inversely proportional to $r^{2 N}$. 

The exterior metric is defined as a pure Lovelock gravity of Lovelock order N in d dimensional spacetime. The metric for exterior spacetime is given by (see some brief recall of pure Lovelock equations in \ref{appendix}) \cite{Gannouji:2019gnb, Dadhich:2015nua, Dadhich:2012cv, Gannouji:2013eka, Dadhich:2015ivt, Padmanabhan:2013xyr, Dadhich:2015lra},
\begin{equation}\label{eqn21}
    ds^2= -f(r) dt^2+ f(r)^{-1} dr ^2 + r^2 d\Omega^2_{d-2};\quad ~\textit{with}~\quad f(r)= 1 - (2^{N}M/r^{d-2 N-1})^{1/N}~,
\end{equation}
where $M=(M^{1/N})^N$ is the ADM mass. Now we consider the interior solution and exterior solution are matches at some radius $R$. Then from \ref{eqn161}, \ref{eqn20}, \ref{eqn21} and the matching condition, we get,
\begin{eqnarray}
B &=&\frac{\left[2 (d-2) k N+ (d-2 N-1)\right]}{(d-2 N -1)}=\frac{1}{\left(1 - (2^{N}M/R^{d-2 N-1})^{1/N}\right)}\nonumber\\
\Rightarrow M^{1/N}&=& \frac{(d-2)k N}{2 (d-2) k N +(d- 2 N -1)} R^{(d-2 N -1)/N}~.\label{eqn22}
\end{eqnarray}
Also, we arrive at the same conclusion by considering $A'$ should be continuous at some radius $R$. Then from the \ref{eqn161}, \ref{eqn19}, \ref{eqn21} and the matching condition, we get,
\begin{eqnarray}
    \frac{A'}{A}&=& \frac{(2 N-d+1) \left( 2^N M R^{2 N-d +1}\right)^{1/N}}{N R \left(\left( 2^N M R^{2 N-d+1}\right)^{1/N}-1\right)}\nonumber\\
    \Rightarrow \frac{2(d-2)k}{R}&=& \frac{(2 N-d+1) \left(2^N M R^{2 N-d+1}\right)^{1/N}}{N R \left(\left(2^N M R^{2 N-d+1}\right)^{1/N}-1\right)}\nonumber\\
    \Rightarrow M^{1/N}&=& \frac{(d-2)k N}{2 (d-2) k N +(d- 2 N -1)} R^{(d-2 N -1)/N}~.\label{eqn24}
\end{eqnarray}
Now we compare this with the Buchdahl limit in Lovelock theory that is $M^{1/N}\leq\frac{2 N (d-N-1)}{ (d-1)^2} R^{(d-2 N -1)/N}$ \cite{Dadhich:2016fku}, and we get the marginal stiffness  to be,
\begin{equation}\label{eqn23}
k= \frac{2 (d-N-1)(d- 2 N- 1)}{(d-2)(d-1)^2- 4 N (d-N-1)(d-2)}~.
\end{equation}
From \ref{eqn23}, it can be easily seen that when $d= 2 N +1$, one will get $k=0$. For $d= 3 N+1$, the $k$ becomes $4/(3 N-1)$. Now for $d=4$, which means $N=1$, we get $k=2$ for that Buchdahl limit for a $d=4$ spacetime that we discuss earlier section. For $d=3N+1$, the \ref{eqn22} or the \ref{eqn24} becomes $\frac{M^{1/N}}{R}= \frac{(3N-1)k}{2 (3N-1)k+1}$. Now for $N=3$, we get $k=1/2$ from \ref{eqn23}. At that value of $k$ and $N$, we get $\frac{M^{1/3}}{R}= \frac{4}{9}$ from \ref{eqn22} or \ref{eqn24}. We can redefine the $M^{1/3}$ as $M$ because $M$ is a constant. Then we get  $M/R= 4/9$. So, we get the possible models of Buchdahl's interior in pure Lovelock gravity at $k=1/2$, which means $P_{t}= \rho/2$, and at the Lovelock order $N=3$.




\section{Discussion }\label{conclusion}

An equation of state,  $p=\frac{1}{2}\rho$ \cite{dad23}, may be present in Buchdahl-like interiors based on recent broad considerations.
However, this is not sustainable on simple physical consideration that at the boundary $p=0$ implies sound velocity being zero, which contradicts $v_s^2 = dp/d\rho =1/2 \neq 0$. One possible way out is to assume $p_r=0$, which takes care of the boundary condition and has $2p_t = \rho$.
In this paper, we have found static fluid solutions of Einstein and pure Lovelock equations with $p_r=0, p_t=k\rho$, which could be a possible model for the interior of a Buchdahl-like star. We have shown that for $k=2$,  we have the possible models of Buchdahl's interior in Einstein's gravity. However, we have shown that for $k=1/2$ and Lovelock order $N=3$, we have the possible models of the Buchdahl interior in pure Lovelock gravity.


\section{Appendix}\label{appendix}
\subsection{Review of Static Spherically Spacetime in Pure Lovelock theory }
Pure Lovelock theory in the D dimension of order N is given by the following Lagrangian \cite{Gannouji:2019gnb, Dadhich:2015nua, Dadhich:2012cv, Gannouji:2013eka, Dadhich:2015ivt, Padmanabhan:2013xyr,Dadhich:2015lra},
\begin{align}\label{LL-lagrangian}
L^{(D)}_{N}=\frac{\alpha_{N}}{16\pi 2^{N}}\delta^{a_{1}b_{1}...a_{N}b_{N}}_{c_{1}d_{1}...c_{N}d_{N}}R^{c_{1}d_{1}}_{a_{1}b_{1}}~...~R^{c_{N}d_{N}}_{a_{N}b_{N}}~.
\end{align}
This theory has the property that even if the Lagrangian is polynomial (of order $N$) in Riemann curvature, the field equations corresponding to the theory are functions of metric and its first two derivatives. 
In particular, the field equations read,
\begin{align}\label{pure_Eq_01}
\mathcal{G}^{a~(N)}_{b}\equiv -\frac{1}{2^{N+1}}\delta ^{ac_{1}d_{1}\ldots c_{N}d_{N}}_{ba_{1}b_{1}\ldots a_{N}b_{N}}R^{a_{1}b_{1}}_{c_{1}d_{1}}\ldots R^{a_{N}b_{N}}_{c_{N}d_{N}}=8\pi \kappa_{(N)} T^{a}_{b}~,
\end{align}
where, $\kappa _{(N)}=\alpha _{N}^{-1}$. Here $ T^{a}_{b}$ is the energy-momentum tensor of the matter fields, which are minimally coupled with $L^{(D)}_{N} $. To solve these equations for a given energy-momentum tensor in a static and spherically symmetric scenario, we take the following metric ansatz,
\begin{align}\label{pure_ansatz}
ds^{2}=-e^{\nu(r)}dt^{2}+e^{\lambda(r)}dr^{2}+r^{2}d\Omega_{D-2}^{2}~.
\end{align}
Since in static and spherically case $\mathcal{G}^{a~(N)}_{b}$ must have only two independent components, we choose the $(t,t)$ and $(r,r)$ components,
\begin{align}\label{pure_Eqtt_02}
\mathcal{G}^{t~(N)}_{t}&=-\frac{1}{2^{N+1}}\left[4N\left(\delta ^{trA_{2}A_{3}\ldots A_{2N}}_{trB_{2}B_{3}\ldots B_{2N}}\right)R^{rB_{2}}_{rA_{2}}\ldots R^{B_{2N-1}B_{2N}}_{A_{2N-1}A_{2N}}+\left(\delta ^{tA_{1}A_{2}\ldots A_{2N-1}A_{2N}}_{tB_{1}B_{2}\ldots B_{2N-1}B_{2N}}\right)R^{B_{1}B_{2}}_{A_{1}A_{2}}\ldots R^{B_{2N-1}B_{2N}}_{A_{2N-1}A_{2N}}\right],
\\
\mathcal{G}^{r~(N)}_{r}&=-\frac{1}{2^{N+1}}\left[4N\left(\delta ^{rtA_{2}A_{3}\ldots A_{2N}}_{rtB_{2}B_{3}\ldots B_{2N}}\right)R^{tB_{2}}_{tA_{2}}\ldots R^{B_{2N-1}B_{2N}}_{A_{2N-1}A_{2N}}+\left(\delta ^{rA_{1}A_{2}\ldots A_{2N-1}A_{2N}}_{rB_{1}B_{2}\ldots B_{2N-1}B_{2N}}\right)R^{B_{1}B_{2}}_{A_{1}A_{2}}\ldots R^{B_{2N-1}B_{2N}}_{A_{2N-1}A_{2N}}\right].
\label{pure_Eqtt_02b}
\end{align}
From the above expressions, one can see that in order to obtain $\mathcal{G}^{t~(N)}_{t}$ and $\mathcal{G}^{r~(N)}_{r}$ for the ansatz \ref{pure_ansatz}, we need to use the following components of the Riemann tensor, 
\begin{align}
R^{tA}_{tB}=\frac{\nu'e^{-\lambda}}{2r}\delta ^{A}_{B}~;\qquad R^{rA}_{rB}=\frac{\lambda'e^{-\lambda}}{2r}\delta ^{A}_{B}~;\qquad R^{AB}_{CD}=\frac{1-e^{-\lambda}}{r^{2}}\delta ^{AB}_{CD}~.
\end{align}
Here we have followed the notation convention as given in the Introduction. Using the above results, we finally obtain
\begin{align}
\mathcal{G}^{t~(N)}_{t}=-\frac{1}{2^{N+1}}\left[\frac{N2^{N}\lambda'e^{-\lambda}(1-e^{-\lambda})^{N-1}}{r^{2N-1}}\Delta _{(2N-1)}+\frac{(1-e^{-\lambda})^{N}2^{N}}{r^{2N}}\Delta _{(2N)}\right],
\end{align}
where,
\begin{align}
\Delta _{(k)}\equiv \delta ^{a_{1}a_{2}\ldots a_{k}}_{b_{1}b_{2}\ldots b_{k}}\delta ^{b_{1}}_{a_{1}}\ldots \delta ^{b_{k}}_{a_{k}}~,
\end{align}
which in $D$ dimension satisfies the recursion relation $\Delta _{(k)}=(D-k-1)\Delta _{(k-1)}$. 
Using this recursion relation, one obtains,
\begin{align}\label{pure_finalEq_03}
\frac{(D-2)!}{(d-2N-1)!}\alpha _{N}\frac{(1-e^{-\lambda})^{N-1}}{2r^{2N}}\Bigg[-Nr\lambda'e^{-\lambda}-(d-2N-1)(1-e^{-\lambda})\Bigg]=8\pi T^{t}_{t}~.
\end{align}
Similarly the  $(r,r)$ component of the field equations leads to, 
\begin{align}
\frac{(D-2)!}{(d-2N-1)!}\alpha _{N}\frac{(1-e^{-\lambda})^{N-1}}{2r^{2N}}\Bigg[Nr\nu'e^{-\lambda}-(d-2N-1)(1-e^{-\lambda})\Bigg]=8\pi T^{r}_{r}~.
\end{align}
In the paper we choose $\alpha _{N}$, such that, $\lbrace (D-2)!/(D-2N-1)!\rbrace 2^{N-2}\alpha _{N}=1$. Now in the case of vacuum spacetime, we get from \ref{pure_finalEq_03},
\begin{align}\label{for-lambda}
\frac{d}{dr}\ln(1-e^{-\lambda})=-\frac{D-2N-1}{N}\frac{d}{dr}\ln r~.
\end{align}
Integrating the above equation and choosing the integration constant such that it corresponds to Schwarzschild spacetime for $D=4$ and $N=1$ leads to,
\begin{align}
e^{-\lambda}= 1- \left(\frac{2^{N}M}{r^{D-2N-1}}\right)^{1/N}~.
\end{align}
Similarly, one can show that,
\begin{align}
e^{\nu}=e^{-\lambda}=1- \left(\frac{2^{N}M}{r^{D-2N-1}}\right)^{1/N}~.
\end{align}
\section*{Acknowledgments}
SB thanks IACS for financial support. CS thanks the Saha Institute of Nuclear Physics (SINP) Kolkata for financial support. SB and CS thank Sumanta Chakraborty  and Naresh Dadhich for various discussions.  CS is thankful to IUCAA, Pune, India, for their warm hospitality and research facilities as the work has been done there during a visit.

\bibliography{mastern}

\bibliographystyle{./utphys1}


\end{document}